\documentclass[journal]{IEEEtran}

\usepackage{graphicx}
\usepackage{subcaption}
\usepackage{upgreek}
\usepackage{amsmath}

\begin{document}

\title{Diffractive coupling for photonic networks:\\ how big can we go?}

\author{Sheler~Maktoobi,
        Luc~Froehly,
        Louis~Andreoli,
        Xavier~Porte,
        Maxime~Jacquot,
        Laurent~Larger,
        and~Daniel~Brunner
\thanks{FEMTO-ST/Optics Dept., UMR CNRS 6174, Univ. Bourgogne Franche-Comt\'{e}, 15B avenue des Montboucons, 25030 Besan\c{c}on Cedex, France} \\e-mail: daniel.brunner@femto-st.fr
\thanks{Manuscript received April 19, 2005; revised August 26, 2015.}}

\markboth{Journal of Selected Topics in Quantum Electronics,~Vol.~, No.~, Month?~2019}%
{Shell \MakeLowercase{\textit{et al.}}: Bare Demo of IEEEtran.cls for IEEE Journals}

\maketitle

\begin{abstract}
Photonic networks are considered a promising substrate for high-performance future computing systems.
 Compared to electronics, photonics has significant advantages for a fully parallel implementation of networks.
 A promising approach for parallel large-scale photonic networks is realizing the connections using diffraction.
 Here, we characterize the scalability of such diffractive coupling in great detail.
 Based on experiments, analytically obtained bounds and numerical simulations considering real-world optical imaging setups, we find that the concept in principle enables networks hosting over a million optical emitters.
 This would be a breakthrough in multiple areas, illustrating a clear path toward future, large scale photonic networks.
\end{abstract}


\IEEEpeerreviewmaketitle

\section{Introduction}
\IEEEPARstart{N}{ETWORKS} are the underlying basis for a large range of physical systems and information processing concepts. 
Particularly in computing, they support a wide range of powerful algorithms such as Hopfield \cite{Hopfield1982} and neural \cite{McCulloch1943} networks as well as in coherent Ising machines \cite{Utsunomiya2011}. 
Especially neural networks have recently resulted in a revolution of modern computing \cite{LeCun2015}. 
One of the fundamental aspects of networks is the propagation of information via \textit{parallel communication} between nodes. 
Parallelism is therefore an essential aspect of networks-based computing, and simultaneously the Achilles' heel in current computing substrates. 

Today, communication over long and intermediate distances relies almost entirely on optical fibers. 
Optical communication is now moving towards shorter ranges, even approaching the level of inter-chip signal transmission. 
Besides the superior energy efficiency and bandwidth, inherent parallelism is a main asset of optical signal transduction. 
This property turned optical neural networks into a field with a long-standing history of interest  \cite{Farhat1985,Denz1998}.
 
Recently, the field has been revitalized by a series of new neural network algorithms \cite{Jaeger2004}, which have resulted in numerous novel demonstrations of neural network computing based on photonic delay systems \cite{Duport2012,Paquot2012,Brunner2013,VanderSande2017,Brunner2018}. 
However, one fundamental property of delay systems is their serial nature, not exploiting the potential parallelism offered by optical processes like diffraction. 
Deep feed-forward neural networks have been realized or discussed using diffraction by complex phase modulations \cite{Lin2018,Chang2018HybridClassification}, or even volume holograms \cite{Wagner1987MultilayerNetworks}. 
We have recently demonstrated the creation of a spatio-temporal photonic reservoir using diffraction \cite{Brunner2015}. 
In our fully-parallel approach, the diffractive coupling concept was leveraged to create a large scale photonic recurrent neural network \cite{Bueno2018}. 
Using a digital micro-mirror device, the photonic network was made trainable based on Boolean connections implemented according to the orientation of micro-mirrors. 
Crucially, the system demonstrated that diffractive coupling is valid for connecting networks hosting hundreds of discrete photonic elements.

In this paper, we investigate fundamental and practical limits to the size of photonic networks coupled via diffraction. 
We consider networks consisting of discrete photonic emitters, the network's nodes, arranged in a periodic array. 
We evaluate coupling of such an node array via a single diffractive optical element (DOE) with a periodic phase-structure, such as in \cite{Brunner2015,Bueno2018}. 
Crucially, we do not require the DOE to be located in the system's Fourier planes.
For infinity-corrected microscope objectives this requirement is hard to fulfill, and we generalized our numerical model to include propagation from and to Fourier planes.
Based on a simple analytical description, we are able to derive the fundamental limit for such a network's size.
This limit is ultimately linked to the paraxial approximation's validity-range.
Outside this range, locations of diffractive-orders and periodically arranged photonic nodes significantly deviate.
Diffractive orders therefore fail to re-inject a photonic node's emission into its neighbors and coupling is lost.
We characterize the practical boundaries based on two independent experiments, each designed to elaborate on fundamentally different limitations. 
Finally, detailed numerical simulations enable clearly identifying the underlying cause and illustrate strategies to go beyond. 
We experimentally confirm coupling for networks hosting up to 30000 photonic nodes. 
Noteworthy, by simply using low magnification microscope objectives with a large numerical aperture (NA), this upper limit can potentially be extended to networks hosting over a million elements, connected fully in parallel. 

The paper is organized as follows: we begin with a general introduction to the diffractive coupling scheme and its application to recurrent neural networks in Section II. 
The concept's state of the art is discussed and a simple yet powerful analytical description which provides the upper limit for diffractive networks' size is given. 
In Section III, we experimentally investigate the concept. 
A single mode optical fiber emulates optical nodes, and coupling for different spatial positions within a network is characterized by translation of the optical fiber's position. 
In a second experiment we emulate this translational effect by tilting the DOE, which allows us to go beyond the practical limitations encountered in the first experimental scheme, and to confirm the analytical limit. 
Finally, Section IV numerically models beam propagation, diffraction and collimation/imaging for an array with an area of 100 mm$^2$.
 
\begin{figure}[t]
\includegraphics[width=0.5\textwidth]{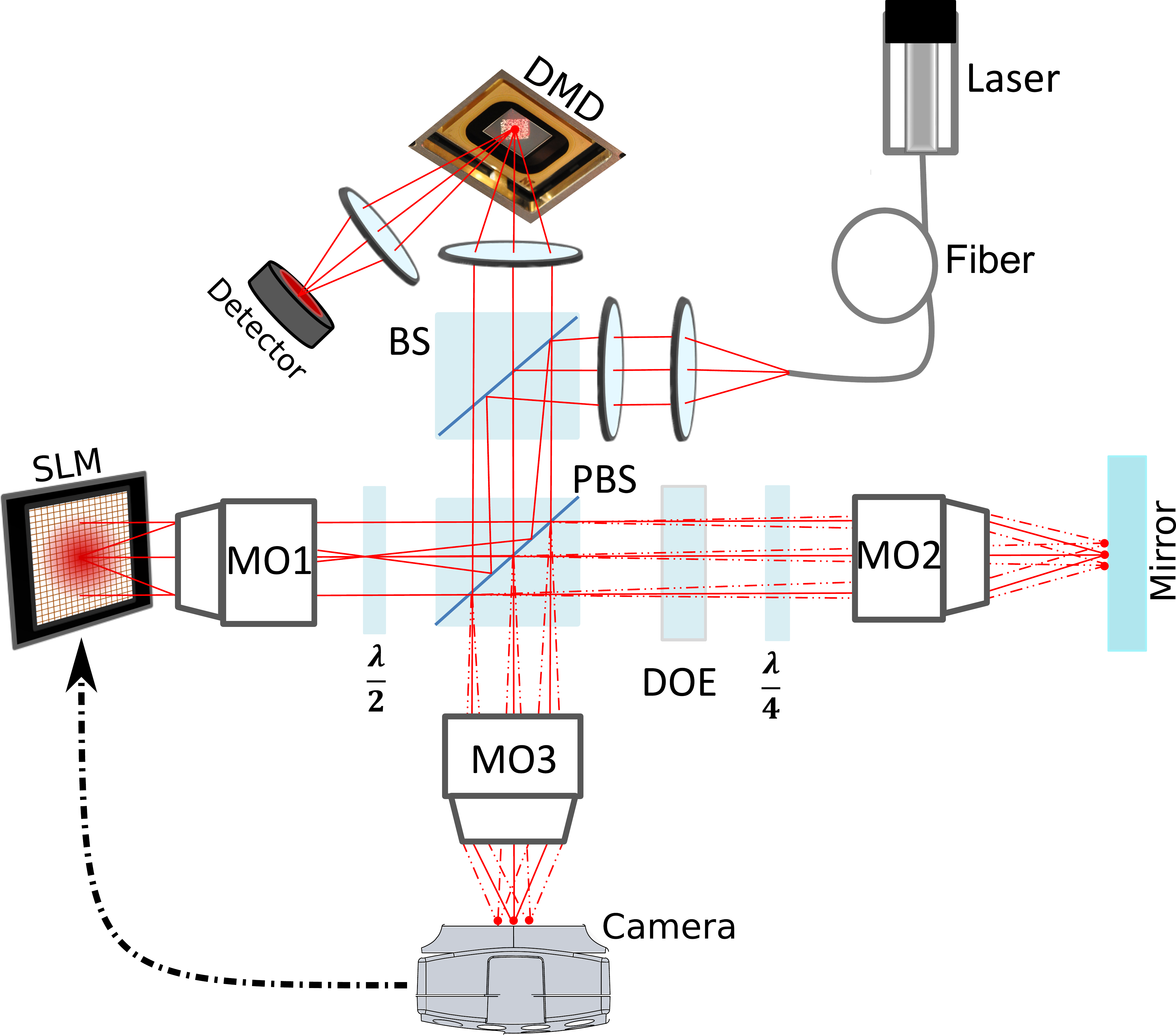}
    \caption{\label{fig:Spatio-temporal}
    Scheme of a diffractively coupled photonic network.
    Network nodes are pixels of a spatial light modulator (SLM), and filtering by a polarizing beam splitter (PBS) brings forward their nonlinearity.
    A diffractive optical element (DOE) located in the beam-path creates coupling between individual photonic nodes.
    Network training is implemented with a digital micro-mirror device (DMD) realizing Boolean selection of SLM-pixels transmitted on the output power meter.
    This system was used to demonstrate learning in the context of a recurrent neural network \cite{Bueno2018}.}
    \vspace{-0.5cm}
\end{figure} 
\vspace{+0.5cm}

\section{Diffractive coupling}

A primary motivation for optical network technology is to establish connectivity fully in parallel for a large number of photonic nodes.
 Nodes are discrete elements arranged into a periodic array with a period $d$.
 Node emission is vertical to the array surface, with $2\Gamma$ and $\lambda$ as their mode field diameter (at $1/e^2$) and emission wavelength, respectively.
 We have previously implemented a photonic network consisting of 900 nodes, using the pixels of a spatial light modulator (SLM) \cite{Bueno2018}, as schematically illustrated in Fig. \ref{fig:Spatio-temporal}.

The SLM is illuminated via a plane wave, and its reflection is filtered by a polarizing beam splitter (PBS).
 Infinity corrected microscope objectives one and two (MO1 and MO2, respectively) image the SLM in the PBS's transmission direction onto the mirror.
 Within this optical path we located a DOE, and each SLM pixel is spatially multiplexed by the DOE's diffractive orders.
 Coupling nodes requires that some of the diffractively multiplexed optical fields overlap.
 The distance $p$ between diffractive orders therefore needs to be close to the array's lattice constant $d$, with a difference $\Delta$ smaller than $\Gamma$.
 This is particularly important for coupling photonic nodes with single-mode emission, i.e. laser arrays \cite{Brunner2015,Heuser2018}.
 In order to enable self-coupling within the network, the overall imaging setup requires a $4f$ architecture.
 In the setup of Fig. \ref{fig:Spatio-temporal}, we therefore included a $\lambda / 4$-waveplate.
 After reflection by the mirror, the optical field double-passes DOE and the $\lambda / 4$-waveplate.
 Due to the resulting 90 degree polarization rotation, the signal is reflected by the PBS and imaged on the camera via MO3.
 The camera's image is used to drive SLM, and hence a recurrent network is established by the closed loop between the SLM and the camera.
 
Besides internal network connections, the system represented in Fig. \ref{fig:Spatio-temporal} and demonstrated in \cite{Bueno2018} was coupled to information injection.
 Furthermore, the signal reflected by the PBS was simultaneously imaged onto the surface of a digital micro-mirror device, which allowed the detection of a weighted network state.
 The weighted state was iteratively adjusted according to a learning rule, and the system corresponds to a large scale photonic neural network which was successfully trained for prediction of the injected chaotic signal.

\begin{figure}[t]
	\includegraphics[width=0.35\textwidth]{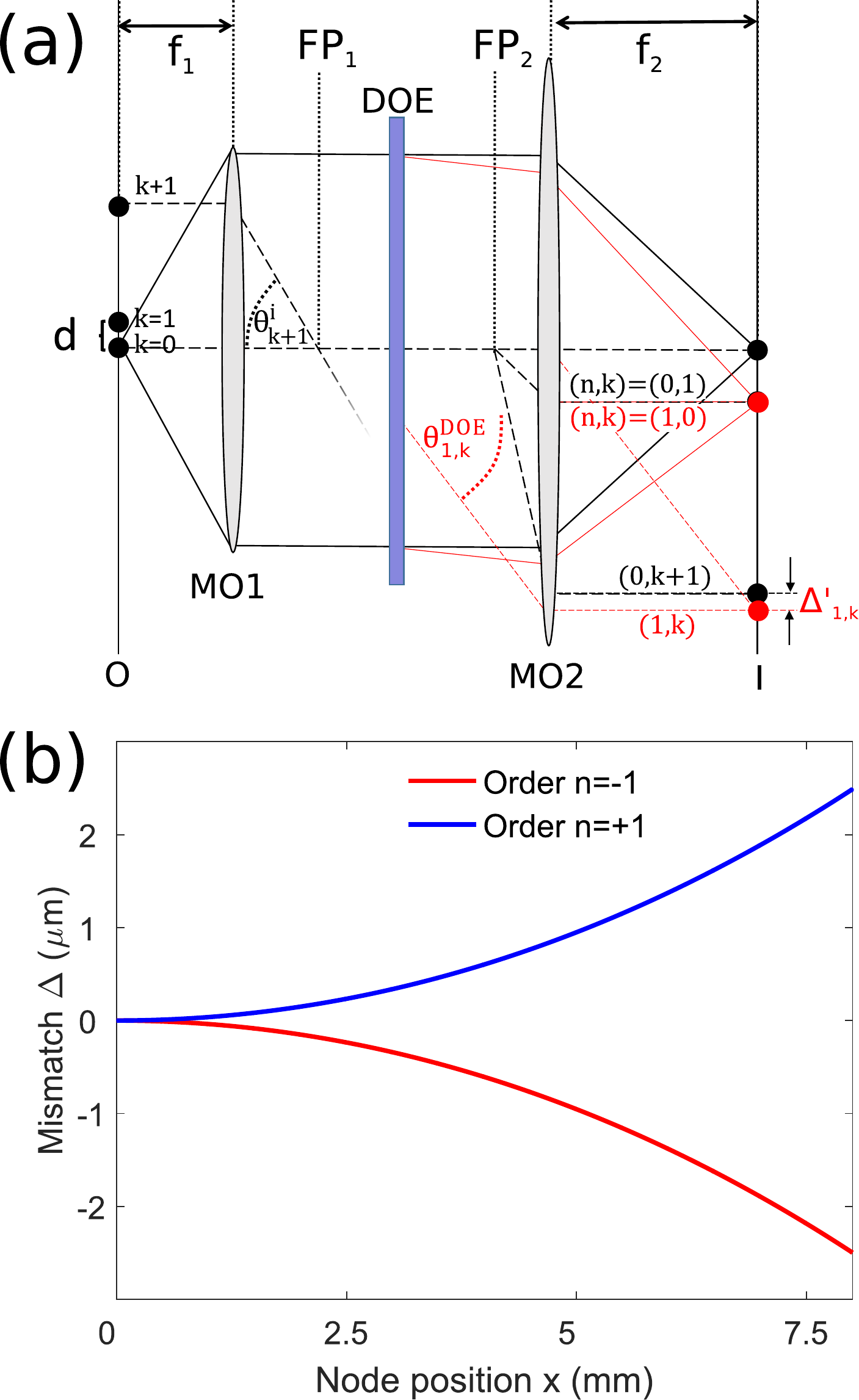}
	\caption{\label{fig:2}
   Scheme to explain the fundamental principle of diffractive coupling.
	(a) Emitters in the plane O are imaged through microscope objectives   MO1, MO2 with respective focal lengths $f_{1}$ and $f_2$ in the plane I.
	The angle of the principle ray of a node with respect to the optical axis is $(\theta_{i})$  and the angle of the +1 diffractive order corresponding to the node is $(\theta_{DOE})$.
	The distance between zero and +1 orders is p and the distance between two nodes is d.
	(b) Mismatch between the positions of nodes and the coupling term due to diffraction versus node positions in plane $O$.}
	\vspace{-0.3cm}
\end{figure}

Here, we explore the principles and limits of diffractive coupling.
 Figure \ref{fig:2}(a) illustrates the functional principle, and elements not essential to this concept have been removed for clarity.
 Optical nodes are periodically arranged at positions $x_k = k \cdot d$ in object plane $O$ and imaged onto image plane $I$, with $k$ being the node index.
 The system's magnification is given by MAG=$f_{2} / f_{1}$, and the position of a node's image is therefore $x_{k}' = \textrm{MAG} \cdot x_{k}$.
 Between MO1 and MO2 we locate the DOE, and in plane $I$ we illustrate nodes and their diffracted images by black and red dots, respectively.
 Infinity-corrected microscope objectives typically have short, in cases even negative back-focal lengths.
 This renders placing the DOE inside Fourier planes FP$_{1}$ and FP$_{2}$ of MO1 and MO2 impractical or impossible.
 We therefore explicitly do not consider a $4f$-system \cite{J.Goodman2017}, a fact which is of particular relevance for numerical simulations.
 
In this setting, coupling nearest neighbors requires that in plane $I$ the optical fields of node $k$'s first diffractive order overlaps with node $k+1$'s zero diffractive order.
 As the DOE is located in the imaging system's collimated space, we can treat its action according to diffraction at infinity, i.e. exclusively considering plane waves.
 The propagation angle $\theta_{k}^{DOE}$ for diffractive order $n$ of node $k$ can be obtained by the grating equation
\begin{equation}\label{eq:GratEq}
sin(\theta_{n,k}^{DOE}) = sin(\theta_{k}^{i}) + n\frac{\lambda}{p^{DOE}},
\end{equation}
\noindent where $\theta_{k}^{i} = \arctan(kd / f_{1})$ is the angle impacting onto the DOE for the collimated beam of node $k$.
 According to Eq. (\ref{eq:GratEq}), we approximate the DOE's action by a periodic diffraction grating with a modulation period of $p^{DOE}$.
 Crucially, imaging a periodic array through a periodic diffraction grating creates spatially multiplexed images of different properties.
 The zero-order image, $n=0$, will reproduce the original array's periodicity.
 However, images formed by other diffractive orders will experience a distortion which continuously increases with larger viewing angles, schematically illustrated in Fig. \ref{fig:2}(a).

In order to capture this effect, we inspect the angular spectrum and the resulting image-positions as a function of $n$ and $k$.
 Focusing by MO2 results in position $\tilde{x}_{n,k}' = f_{2}\cdot\tan(\theta_{n,k}^{DOE})$.
 This needs to match the position of node $(k+n)$'s zero order image, and hence we define the mismatch between both as $\Delta' = | x'_{0,k+n} -  \tilde{x}'_{n,k}|$.
 Using $\tan(\theta) = \sin(\theta) / (\sqrt{1 - \sin(\theta)^2)}$ and $\sin(\theta^{i}_{k}) = k d / \sqrt{f_{1}^2 + k^2  d^2}$, we obtain
\begin{equation} \label{eq:Delta}
\Delta_{n,k}' =  \frac{f_{2}}{f_{1}}(k+n)d - \frac{\left(\frac{k d}{\sqrt{{f_1}^2+k^2  d^2}} + n\frac{\lambda}{p^{DOE}}\right)f_2}{\sqrt{1-\left(\frac{k d}{\sqrt{{f_1}^2+k^2 d^2}} + n\frac{\lambda}{p^{DOE}}\right)^2}}.
\end{equation}\\
\noindent This equation corresponds to a simplified analytical model to explore the diffractive coupling concept's limits.
 It only considers the effects of diffraction and imaging, assumes that all optical elements are thin, not separated by a physical distance (effects of propagation are not included) and free of aberrations.
 Finally, we scale the mismatch by the imaging system's magnification to obtain $\Delta_{n,k} = \Delta_{n,k}' / \textrm{MAG}$.

According to Eq. \ref{eq:Delta}, coupling is only established for correspondingly adjusted values of $p^{DOE}$, $d$ and $\lambda$.
 Generally, we impose $\Delta=0|_{k=0}$, hence optimize coupling for the array's center.
 This leads to
\begin{align}
 \theta^{i}_{1} &= \arctan(\frac{d}{f_{1}}) , \label{eq:AlignImage} \\
 \theta^{DOE}_{1,0} &= \arcsin(\frac{\lambda}{p^{DOE}}) , \label{eq:AlignDOE} \\
 \theta^{DOE}_{1,0} &= \theta^{i}_{1} \label{eq:Align} .
\end{align} 
\noindent as the alignment condition for the system.
 In the experiment shown in Fig. \ref{fig:Spatio-temporal}, $p^{DOE}$ and $d$ are predetermined, which leaves wavelength $\lambda$ as a free parameter.
 As the SLM hosting the photonic nodes is a broad-band device, the choice of $\lambda$ is sufficiently flexible to select a commercial DOE which satisfies coupling for the spacing between SLM pixels.
 In other experimental setting this might not be the case.
 In particular for arrays with active photonic nodes, $\lambda$ may not be adjustable for satisfying Eq. \ref{eq:Align}.
 There, one requires to either specifically tailor pitch $d$ or $p^{DOE}$.

\section{Experimental evaluation}

\begin{figure}[t]
\includegraphics[width=0.45\textwidth]{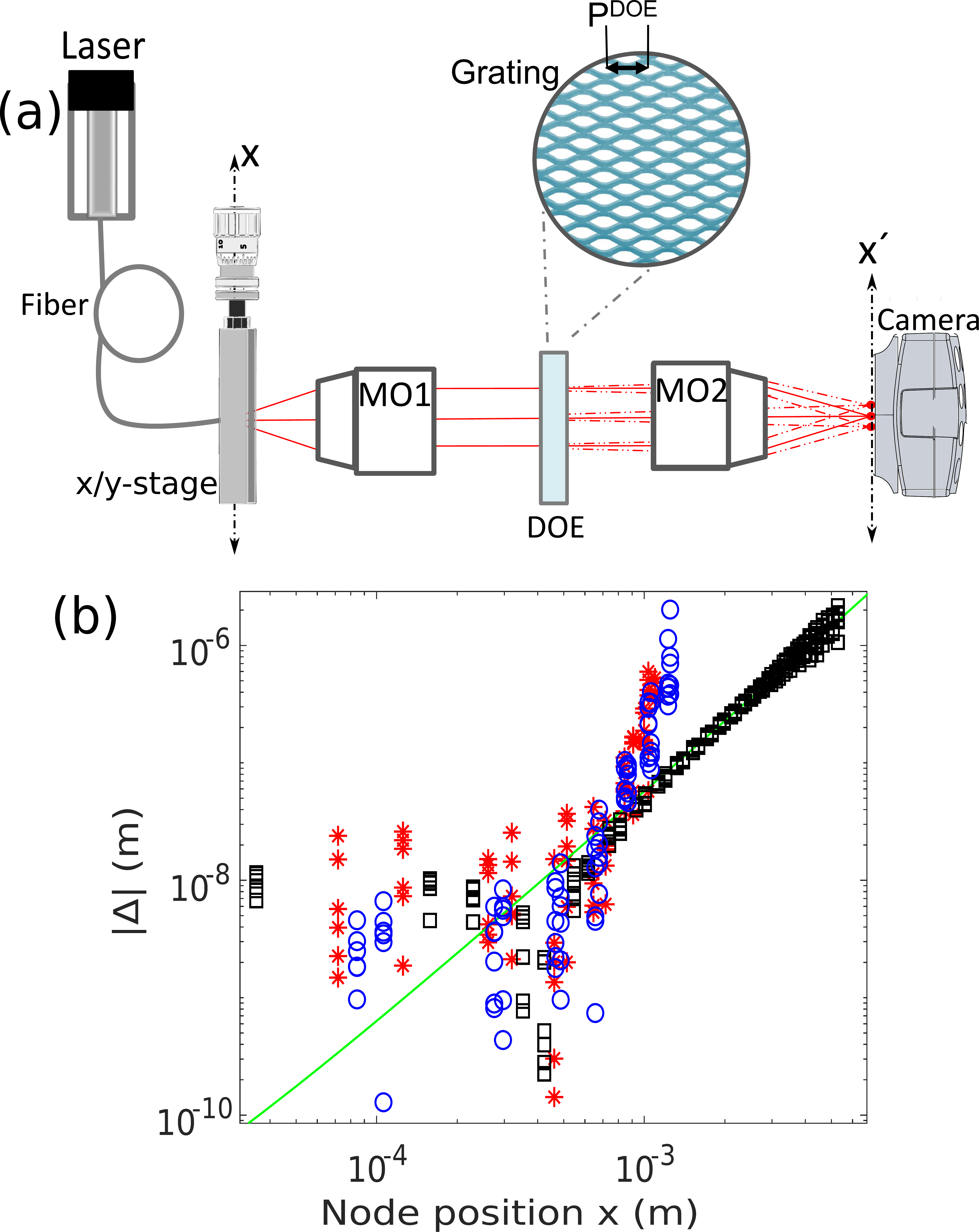}
    \caption{\label{fig:exp3}
    (a) Experimental setup testing diffractive coupling limits by translating a single-mode fiber across the object plane. 
    (b) Mismatch $| \Delta |$ identifies the validity limit of diffractive coupling.
    Equation \ref{eq:Delta} (green line) illustrates the fundamental limit. 
    Experimental (red stars) and numerical results (blue circles) excellently agree (NA1=0.25, NA2=0.1).
    The black squares numerically demonstrate that for NA1=0.45, NA2=0.2 the system would approach the analytical limit.}
    \label{fig:exp1}
\end{figure}

We emulate the original array of photonic nodes by translating a single mode source along the (x,y)-plane.
 We precisely measure the displacement of the diffraction maxima and study the limits posed by fundamental mechanisms of diffractive coupling, and by limitations of our experimental system.
 The experiment is schematically illustrated in Fig.~\ref{fig:exp1}(a).
 This particular setup emulates the relevant physical mechanisms of a photonic network, such as illustrated in Fig. \ref{fig:Spatio-temporal}.
 The point source impersonating the photonic nodes is the tip of a single-mode optical fiber mounted on a x/y-stage (Thorlabs ST1XY-D/M), which gives us micrometer-precision control of its displacement.
 The fiber guides the emission of a laser diode (Thorlabs LP660 SF20, $\lambda = 662.1$~nm) and has a mode field diameter at its output of $2\Gamma=(4\pm 0.5)~\mu$m.
 Its output tip is placed at the focal plane of a microscope objective (MO1: Olympus RMS10X, $f_1$ = 18~mm, NA1=0.25).
 After the DOE (HOLOOR MS-443-650-Y-X, $p^{DOE}\sim1.3~$mm), a second microscope objective (MO2: Olympus RMS4X, $f_2$ = 45~mm, NA2=0.1) focuses the diffracted beams.
 Thus, the emission emulating a photonic neuron is uniformly distributed along an array of (3$\times$3) diffractive orders, which are recorded with a CMOS camera (IDS U3-3482LE-M).

Figure~\ref{fig:exp1}(b) shows coupling mismatch $| \Delta |$ versus a node's position $x$ over two orders of magnitude and on a double-logarithmic scale.
 Experimental data was obtained from fitting the image recorded by the camera with an array of nine Gaussian modes, whose center-positions are used to calculate mismatch $\Delta$
 Analytically calculated data (green line) are compared to experimentally measured mismatches (red stars).
 While the analytic dependence shows a clear polynomial increase of mismatch $\Delta$, we observe a number of unclear features in the experimental data.
 For node displacements below $0.5~$mm, experimental data is dispersed and does not follow a clear trend.
 For displacements in this range the expected $\Delta$ is blow 100 nm.
 Detection noise of the camera and a limited precision of the fitting routine therefore result in uncertainties larger than the expected $| \Delta |$.

Experimental and analytical results agree well in the range of $0.1~$mm $<x<0.5~$mm.
 However, for experimental data obtained with a node displacements of $x>1~$mm, we observe a strong divergence with respect to the analytical law.
 We postulate this divergence is due to a beam-vignetting effect caused by a too small NA of the first microscope objective (NA1).
 As the photonic node moves out of the object plane's center, the outer-parts of its emission-cone leave MO1's clear aperture.
 Detrimental diffraction at the clear aperture's edge is the consequence, inducing additional aberrations.
 We test this hypothesis by numerical simulations of the experimental setup, which are discussed in Section IV.

\begin{figure}[t]
	\includegraphics[width=0.45\textwidth]{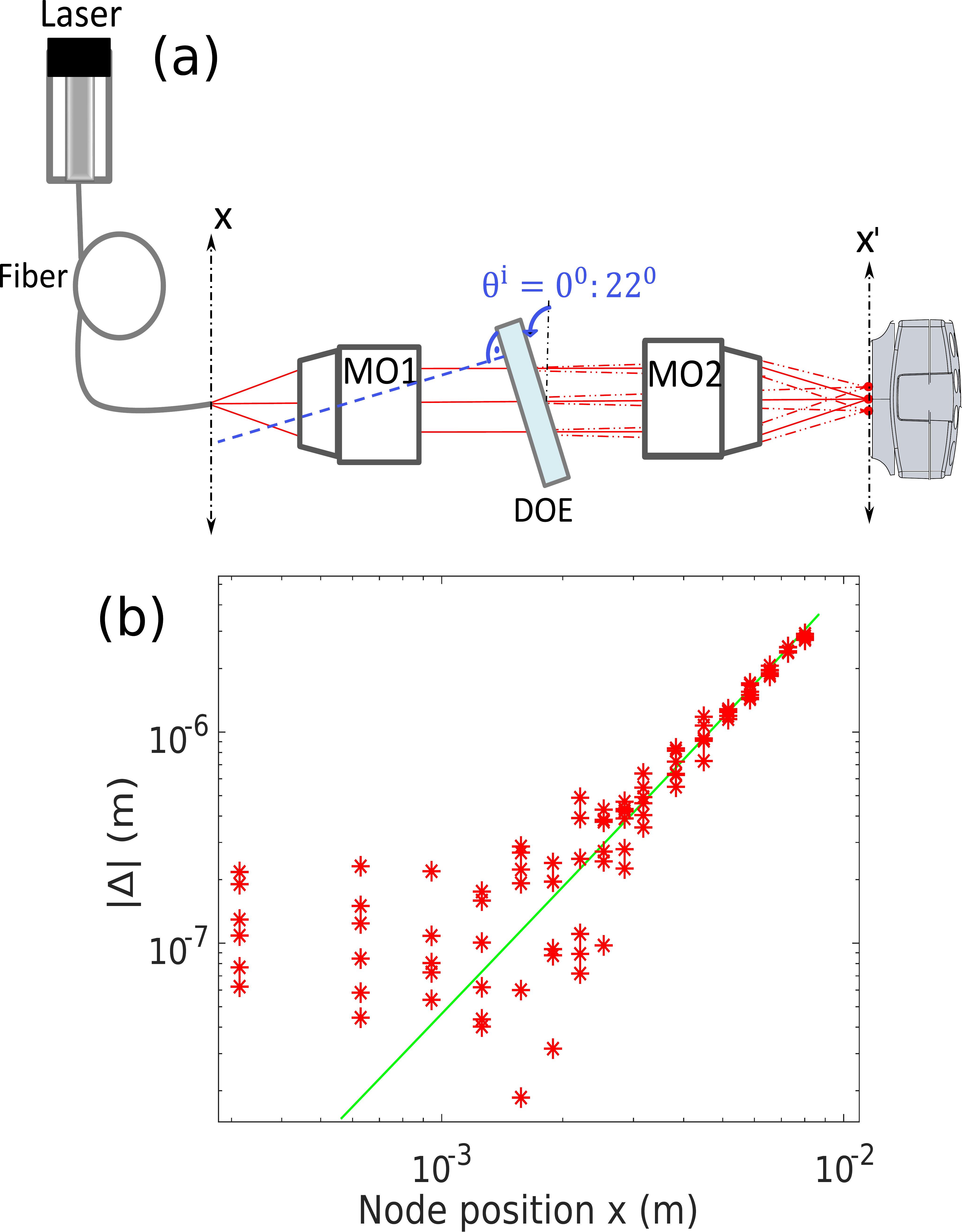}
	\caption{\label{fig:4}
		(a) Experimental scheme to test limits of diffractive coupling by tilting DOE relative to the optical system's central axis.
		The motivation is to emulate positions 0~mm$\leq x \leq$8~mm without causing beam-vignetting by the microscope objectives. 
		(b) The diffractive coupling mismatch $| \Delta |$ obtained by the experiment (red stars) and the analytical limit (Eq. \ref{eq:Delta}) excellently agree.}
\end{figure}

From the previous results we can conclude that the size of a photonic network would not be limited by the diffractive coupling concept per se.
 In a second experiment, we exclude possible limitations caused exclusively by imaging nodes at large displacements in the (x,y)-plane.
 Instead of translating the fiber tip we therefore tilt the DOE relative to the now continuously centered, collimated beam. 
 Figure~\ref{fig:4}(a) illustrates this experimental approach. 
 There, DOE-tilt angle $\theta^{i}$ emulates translation of the fiber tip according to $x = f_1 \cdot \tan(\theta_{i})$.
 In the experiment we used $f_1=20$ mm (MO1: Nikon N10X-PF).

Figure~\ref{fig:4}(b) illustrates the distance mismatch for tilt angles $\theta^i$ ranging from 0 to 22 degrees, which is equivalent to emulating photonic neuron positions $x$ ranging from 0 to 8 mm. 
 This isolates the effects of a node's position onto diffractive coupling, and we obtain an excellent agreement between the analytical model (green line) and the experimental data (red stars). 
 Results confirm the validity of diffractive coupling far beyond what is currently demonstrated. 

\begin{figure}[t]
	\includegraphics[width=0.45\textwidth]{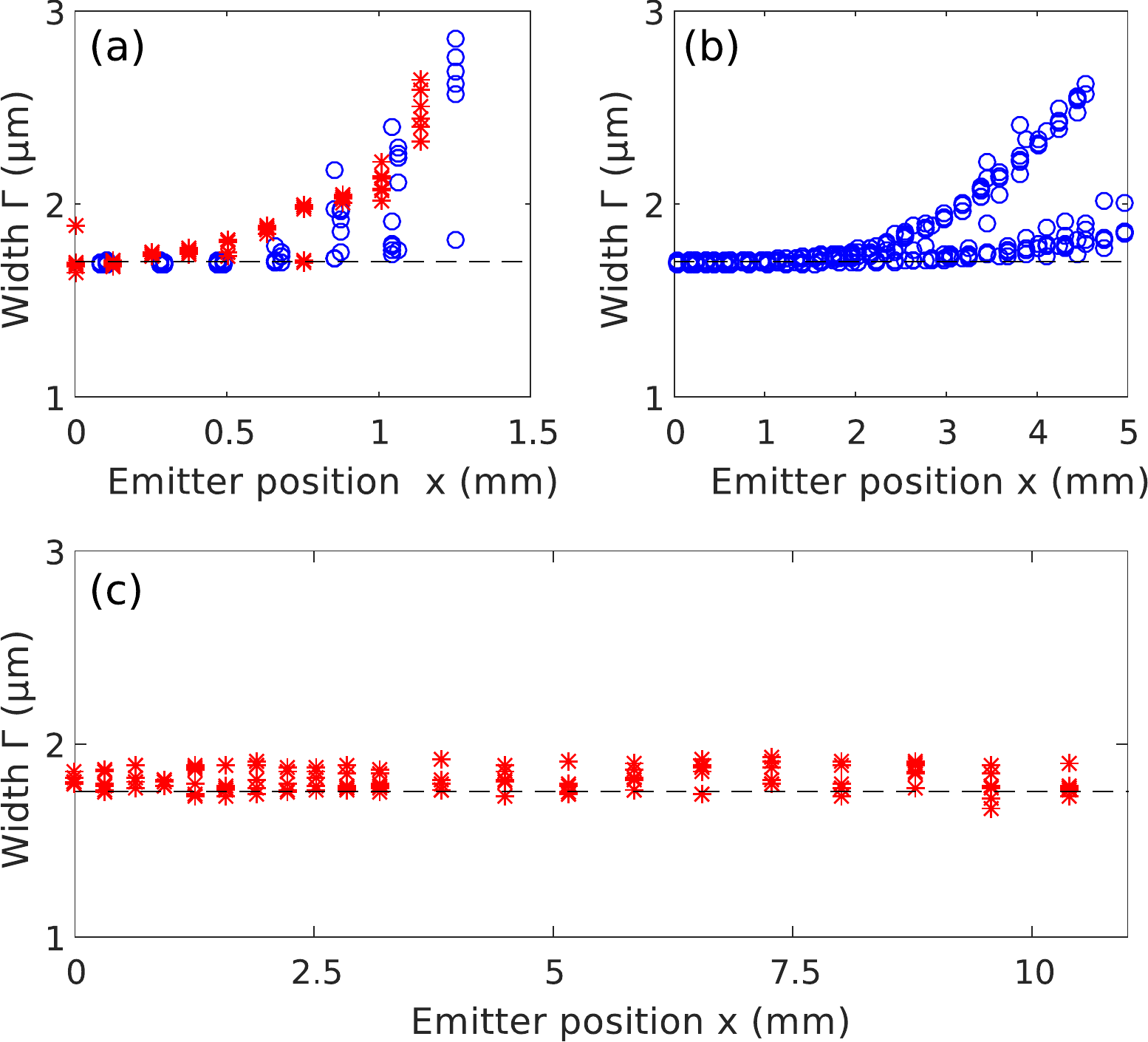}
	\caption{\label{fig:exp5}
		Width of the diffractive orders (red stars) relative to the diffraction limit (dashed line), and comparison to numerical simulations (blue circles).
		(a) Experimental data obtained with the setup shown in Fig. \ref{fig:exp1}(a).
		For NA1=0.25, vignetting leads to a rapid increase of the width $\Gamma$ for $x > 0.5$mm.
		(b) Numerical simulations based on NA1=0.45 indicate the potential for significant improvements.
		Data in panel (c) was obtained with the setup shown in Fig. \ref{fig:4}(a).
		Diffractive coupling does therefore not significantly influence the system's optical resolution.
		Comparison between (b) and (c) illustrates the high sensitive of the systems resolution to even small vignetting.}
	\vspace{-0.3cm}
\end{figure}

However, establishing high-quality optical coupling between the elements of a periodic photonic node array also depends of the system's points spread function, i.e. its optical resolution. 
 Figure~\ref{fig:exp5} demonstrates the impact of various effects upon width $\Gamma$ obtained from fitting the diffractive orders in image plane $I$. 
 The all-important diffraction limit is illustrated by the dashed line.
 Figure~\ref{fig:exp5}(a) characterizes $\Gamma$ of a single-mode photonic node for different positions $x$ in the object plane $O$, showing red stars (blue circles) as the experimental (numerical) data for NA1=0.25 and NA2=0.1.
 As previously reported for mismatch $\Delta$, width $\Gamma$ also diverges from the fundamental limit for node positions beyond $x=0.8~$mm. 
 This similarity suggests beam vignetting as the underlying reason, which we confirm by numerical simulations using the large NA setting (NA1=0.45, NA2=0.2), Fig.~\ref{fig:exp5}(b).
 The MOs' larger collection angle significantly increases the vignetting-free zone in the object plane.
 Finally, we turn to the experimental setting of Fig.~\ref{fig:4}(a), and emulate different object-plane positions by tilting the DOE.
 Experimental data is shown as red stars Fig.~\ref{fig:exp5}(c).
 As vignetting is avoided, $\Gamma$ does not exhibit any significant broadening for DOE-tilt angles corresponding extreme positions $x>10~$mm.
 We would like to point out that $x>10~$mm corresponds to DOE tilt-angles beyond 20$^{\circ}$, and while non-paraxial contributions fundamentally limit the concept through $| \Delta|$, they do not affect the system's optical  resolution.
 However, what our results illustrate is the concept's sensitivity to general limitations of the imaging system.
 Considering an array with a radius of 2~mm and nodes spaced by 10 $\mu$m, our experiments confirm the potential for creating photonic networks hosting 30.000 nodes with our current experimental setup.
 In this entire region width $\Gamma$ remains close to the resolution limited.

\vspace{+0.5cm}
\section{Numerical method}

In the previous section we observed effects in experiments (i.e. strong divergence of $\Delta$ and increase of $\Gamma$ due to vignetting) which cannot be accurately captured with the simple analytical treatment of Section II.
 A correct description of the underlying optical processes requires considering effects of coherent light propagation through the imaging system.
 Crucially, we cannot make use of the paraxial approximation as it imposes $\sin(\theta) = \tan(\theta)\approx\Theta$, which in turn leads to $\Delta=0$ for any $x_k$.
 Two main propagation algorithms are used.

First, we use the non-paraxial plane wave spectrum approach between MO1 and MO2.
 This is a common method for describing propagation of an electromagnetic field with a particular spatial amplitude and phase distribution \cite{J.Goodman2017}.
 Propagating the optical field from one plane to the next utilizes the field's fast Fourier Transformation ($FT$), resulting the fields planar wave spectrum $A(\nu_{x},\nu_{y})$ .
 The $FT$ creates a decomposition of the incident field into complex-valued plane waves located in spatial frequency space with dimensions $\nu_{x}$ and $\nu_{y}$.
 For each resulting plane-wave component propagation simply corresponds to translating their phase determined by the distance to be propagated.
 Crucially, phase-translation $\rho$ for each planar wave component is calculated without employing approximations $\tan(\theta)= \sin(\theta)\approx\Theta$ and $\cos(\theta) \approx 1-\theta^{2} / 2$, resulting in $\rho = \sqrt{1/\lambda^2 - \nu_{x}^{2}- \nu_{y}^{2} }$.
 Finally, with an inverse Fourier transform of the such propagated plane wave components, one obtains the final optical field at new position $(x,y,z)$.
 Propagation between optical elements along the z-direction is computed in the $(x,y)$-plane according to
\begin{equation} \label{eu_eqn}
A(x,y,z)=FT\left\lbrace A(\nu_{x},\nu_{y})e^{i2{\pi}z\rho} \right\rbrace .
\end{equation}
 To include the effect of the DOE we modulated the optical beam between both microscope objectives according to the DOE's phase profile. 
 The phase profile was obtained directly from the component based on a phase retrieval technique \cite{shechtman2015}.
 Based on the such obtained characterization we approximated the DOE by the likely assumption that the DOE implement a binary phase modulation.

Second, we employ the Debye integral method to simulate the propagation through high NA microscope objectives \cite{born2013,wolf1981,sheppard2000,Leutenegger2006}.
 Propagation of an optical beam with $\lambda=660~$nm and a radius of 5 mm through a microscope objective with a focal distance of 18 mm (i.e. Olympus RMS10X) creates of the order of $\sim 2\cdot 10^3$ phase-slips across each direction of the (x,y)-plane.
 Considering the Nyquist sampling criteria, this requires sampling the wave's phase profile with at least ($4\cdot 10^{3}\times4\cdot 10^{3}$) data points.
 Arrays of such size render the use of a standard desktop PC unpractical.
 As can be seen in Fig. \ref{fig:exp3}(b), our characterization is highly sensitive to artifacts, which would most likely results in an even larger sampling matrix.

In the Debye-method, optical field components are projected onto the microscope objective principal planes, one of which is a sphere centered at the image's focal point.
 This approach does not employ the scalar field simplification and is valid if the differential phase-change with respect to the principal plane is small.
 The field in the image plane is closely approximated by the Fourier transform of the such projected incident field.
 We therefore first project the incident field onto cylindrical coordinates $(r,\theta, \phi)$, decomposing the optical field into its radial (p-polarized) and tangential (s-polarized) components.
 The transformed field is then projected onto the microscope objective's principle plane in spherical coordinates.
 The resulting electric field $\vec{E}$ at a point $(x,y,z)$ near the focus is obtained by integrating the propagated plane waves \cite{Leutenegger2006},
\begin{equation}\label{eu_eqn1}
\begin{split}
E(x,y,z)= \frac{-if}{\lambda_{0}}\int_{0}^{\pi}sin(\theta)\cdot \\
\int_{0}^{2\pi}   E_{t}(\theta,\varphi) e^{ i(k_{z}z-k_{x}x-k_{y}y)} d\theta d\varphi . \end{split}
\end{equation}
\noindent Equation \ref{eu_eqn1} can be rearranged using the Fourier transformation, which results in
\begin{equation} \label{eu_eqn2}
E(x,y,z)= \frac{-if}{\lambda_{0} {k_{t}^{2}}} FT \left\lbrace\frac{E_{t}(\theta,\varphi) e^{ i(k_{z}z)}}{cos\theta }\right\rbrace.
\end{equation}

However, the field-distribution calculated by Eq. (\ref{eu_eqn2}) must be corrected, as the result is only correct inside a lateral surface of the order of $\left({\frac{\lambda}{2\cdot NA}}\right)^{2}$.
 This criteria is equivalent to a Fresnel approximation.
 In the frame of our work we require the numerical model to be valid outside this limit, as the objective is to simulate networks of a size exceeding 1 mm.
 We therefore implement a treatment with larger generality, which enables the extension of the Debye-method to ranges outside the focal volume.
 The principle is a coordinate-rescaling which accounts for the relation between spatial positions $(x,y)$ and spatial frequencies $(\nu_{x},\nu_{y})$ in an imaging system beyond the paraxial approximation.
 In the paraxial approximation, the relation between image plane and the spatial frequency coordinates are simply $x_{0}|y_{0}=f\cdot\lambda\cdot\nu_{x|y}$, whereas in the non-paraxial case the relation is   $x|y=(f\cdot\lambda\cdot\nu_{x|y}) / {(\sqrt{1-\lambda^{2}\cdot(\nu_{x|y})^{2} }}$.
 Hence, both coordinate systems are connected via $x|y=(f\cdot x_{0}|y_{0}) / (\sqrt{f^{2}-(x_{0}|y_{0})^{2}})$, which is the rescaling which we applied to all numerical simulations.
 The results are in remarkable agreement with the experiment.

Both techniques are combined in a sequence in order to describe our optical system.
 First, we use the inverse Debye-method to calculate the optical field in FP1.
 From there, we propagate the field to the DOE, apply the DOE's phase modulation, propagate to FP2 and finally calculate the optical field in image plane $I$ via the Debye-method. 
 Results of the numerical simulations are shown in Fig. \ref{fig:exp3}(b).
 Here, we simulated an area of 25 mm$^2$ with an overall number of $(10^{12}\times 10^{12})$ samples.
 We accurately reproduced the experimental setup, including the directly measured distances from the MO1 to the DOE and from the DOE to the MO2, plus the physical properties of both microscope objectives as obtained from their data sheets.
 Results are shown as blue circles, and they agree with the experimental findings (red stars) exceptionally well, both, qualitatively as well as quantitatively.
 Crucially, we used the same fitting and data-extraction routing as for the experiment.
 The dispersion of $\Delta$ for $x$ smaller than $\sim0.5~$mm is consequence of the fitting routine's uncertainty and of the limited spatial resolution.

The confirmed divergence between our experimental results and the analytical limit through the aberration-free numerical imaging system suggests beam vignetting as the underlying cause.
 We therefore modified the numerical simulation and replaced MO1 with a high NA, low magnification microscope objective.
 Crucially, such devices are available off-the-shelf.
 As we can see, for this configuration the numerical simulation collapses on top of the analytical upper limit of the system's size.
 This is an excellent results, as it confirms the validity of diffractive coupling under conditions comparable to a realistic experimental setting.
 Crucially, data of Fig. \ref{fig:exp3}(b) suggests that for optimized microscope objectives the mismatch $|\Delta|$ remains below $1~\mu$m when coupling nodes in an array of a radius of 4 mm.
 Considering a realistic spacing of $10~\mu$m between photonic nodes, this confirms the diffractive coupling concepts for networks hosting \textit{over a million} optical nodes.

\vspace{+0.5cm}

\section{Discussion}

Our experimental, numerical and analytical investigation provides the first systematic analysis of size-limits for optical networks coupled via diffraction.
 The here employed experiments only serve the identification of different limiting effects, and other considerations have to be taken into account when coupling real networks.
 A vitally important parameter is the distance between microscope objectives, as too long distances results in very strong beam-vignetting at MO2.
 The same consideration prohibits increasing NA1 through a high magnification MO1, as this results in a steepening relationship between $x_{k}$ and $\Theta_{k}^{i}$, again amplifying beam-vignetting at MO2.
 Finally, in an all optical network, for using arrays of semiconductor lasers \cite{Brunner2015,Heuser2018}, coupling takes place in the object plane.
 This results in $MAG=1$, and the second microscope objective can be ignored as long as vignetting is avoided.
 Another effect to consider is that the DOE's highest spatial frequency will be the leading term in $\Delta$.
 Divergence will therefore be amplified using DOE's to realize longer-range as well as more complex than next-neighbor coupling, as well as in systems where array-pitch $d$ is large.

One could argue that the breakdown of diffractive coupling due to $\Delta$ can be interpreted as an aberration, and hence aberration-management could be employed to increase the size of potential networks even further.
 However, according to Eq.~\ref{eq:Delta} and as illustrated by Fig. \ref{fig:2}(b), the signs of diffractive order $n$ and of displacement $\Delta$ are related.
 When interpreted in terms of optical distortions, then $n>0$ results in pincushion-type distortions, while $n<0$ leads to barrel-type distortions.
 Correctional optics for compensating $\Delta$ therefore would have to compensate for both effects at the same position $(x,y)$ and for very similar $(\nu_x,\nu_y)$ simultaneously.
 We have for now not been able to identify a mechanism capable of devising corrections to beams which, after passing the DOE, are nearly degenerate.
 It appears that the here derived limit is a fundamental hurdle.

\vspace{+0.5cm}

\section{Conclusion}
In our work we have analyzed the validity range for diffractive coupling with great detail.
 Mismatch $\Delta$ determines the mismatch between a diffractive order's ideal and real position, and we determine its dependency on the node's position in two complementary experiments.

By translation of a single-mode fiber's output, the experiment emulates different node positions and is sensitive to aberrations as they would be found in an optical setup for photonic networks.
 Mismatch $\Delta$ quickly increases in polynomial fashion for nodes located further than $\approx0.7$ mm away from the array's center.
 Crucially, the found dependency does not agree with the fundamental limit of diffractive coupling, and we identify beam vignetting as the underlying cause.
 Despite this limitation, the setup is well suited for coupling up to 30000 photonic nodes for a spacing of $d=10~\mu$m.
 In a second experiment we emulate diffraction of nodes at different positions by tilting the DOE.
 This removes the angled propagation of the system's zero order, hence  isolates fundamental coupling limitations from the vignetting effect. 
 We are able to demonstrate excellent agreement between the experimental and analytical description.

Based on this insight, we modify system parameters in the numerical simulation and simulated a low magnification, high NA microscope objective (NA1=0.45).
 Here, beam vignetting is strongly reduced and mismatch $\Delta$ perfectly agrees with the analytically derived limit for nodes located up to 6 mm away from the array's center.
 Importantly, mismatch $\Delta$ remains below 1$~\mu$m within an area of $\sim 50~$mm$^2$, which for the size of typical optical nodes such as SLM pixels or laser's could be seen as a limit for coupling.
 We can therefore estimate an upper bound of diffractively coupling over 1 million nodes located in an array spaced by $d=10~\mu$m.
 
We have adjusted the numerical model such that it does not make use of the typically employed approximations.
 Most notably, we do not employ the paraxial beam approximation during the propagation of plane wave.
 Furthermore, we have introduced a rescaling of the image plane.
 This novelty was made necessary by the use of the Debye integral when calculating the action of the microscope objectives.
 The phase profile of the DOE was obtained by a phase-retrieval method, and hence our simulation represents an excellent description of the diffractively multiplexed imaging system.

\appendices

\section*{Acknowledgment}
This work was supported by the EUR EIPHI program (Contract No. ANR-17-EURE-0002), by the BiPhoProc ANR project (No. ANR-14-OHRI-0002-02), by the Volkswagen Foundation NeuroQNet project and the ENERGETIC project of Bourgogne Franche-Comt\'{e}.
 X.P. has received funding from the European Union’s Horizon 2020 research and innovation programme under the Marie Sklodowska-Curie grant agreement No. 713694 (MULTIPLY).

\bibliographystyle{IEEEtran}  
\bibliography{references} 

\begin{IEEEbiography}[{\includegraphics[width=1in,height=1.25in,clip,keepaspectratio]{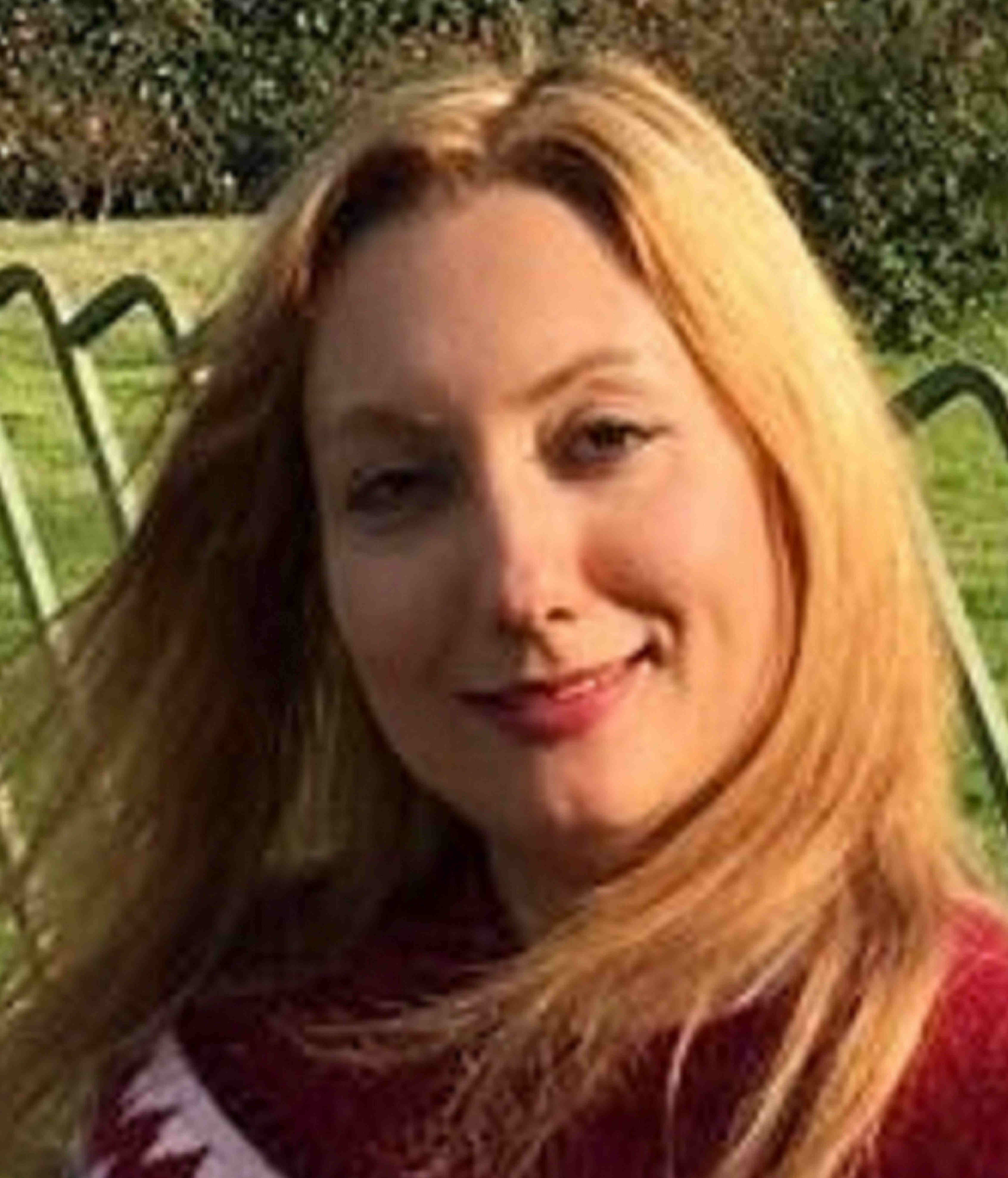}}]{Sheler Maktoobi}
Sheler Maktoobi received the Master degree in 2015 in telecommunication. Since 2016, she has been working toward the Ph.D. degree in photonic neural networks at Femto-st institute for the OPTICS department, University of Bourgogne Franche-Comt\'{e}, Besan\c{c}on, France.
She received the best poster prize of "6th International Symposium in Optics and its Applications", Trento(Italy) and second prize for "ActInSpace competition", Besan\c{c}on (France) in 2018.
Her research interests include optical networks, hardware neural networks, diffractive coupling analysis, modeling, and numerical simulations.
\end{IEEEbiography}

\begin{IEEEbiography}[{\includegraphics[width=1in,height=1.25in,clip,keepaspectratio]{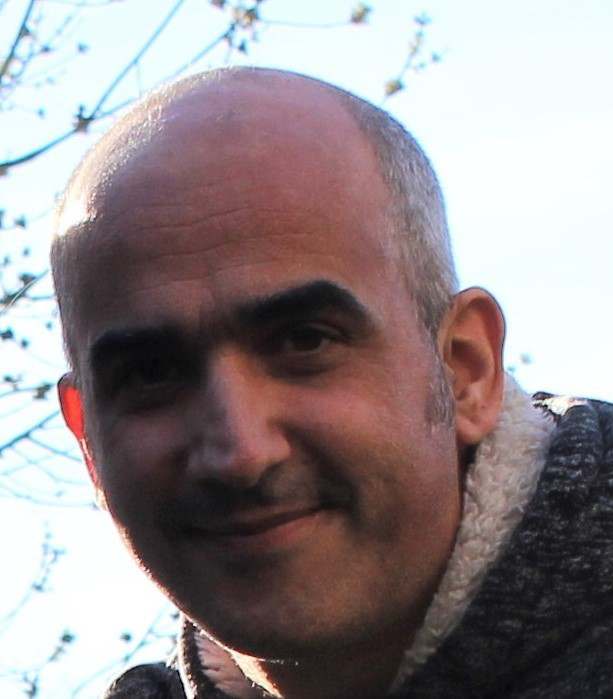}}]{Luc Froehly}
Luc Froehly was born in 1972 in Besan\c{c}on (France), he obtained his PhD in photonics on refractive index metrology in heterogeneous mediums in 2000. From 2001 to 2004 he was a research and teaching assistant at the Swiss Federal Institute of Technology in Lausanne, Switzerland (EPFL). He then developed an activity on Optical Coherence Tomography. At the end of 2004 he became Research Fellow at the CNRS in the Optics Department of the FEMTO-ST Institute.  It continues to develop its optical imaging activities until 2010. Since 2010, his activities have focused more on metrology and beam shaping, in the context of high aspect ratio laser ablation (activity led by François Courvoisier). Recently he also has a connection with photonic neural networks as part of Daniel Brunner's activity. The common denominator of all these activities is the so-called Coherent photonics, which is his field of expertise. 
\end{IEEEbiography}

\begin{IEEEbiography}[{\includegraphics[width=1in,height=1.25in,clip,keepaspectratio]{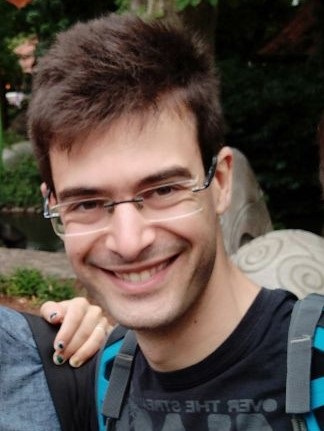}}]{Louis Andreoli}
	Louis Andreoli was born in Besan\c{c}on, France, in 1992. He received the Bachelor’s degree in Physics from Universite de Bordeaux and the Master’s degree in Photonics and Applied Physics from Universite de Franche-Comte. Since 2016 he is a Ph.D. student at the FEMTO-ST Institute, Besan\c{c}on. 
	The main objective of his thesis is the hardware implementation of parallel photonic spatio-temporal neural networks. His research interests include optics, complex system and nonlinear dynamics.
\end{IEEEbiography}

\begin{IEEEbiography}[{\includegraphics[width=1in,height=1.25in,clip,keepaspectratio]{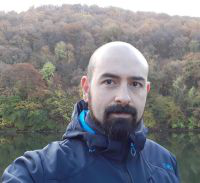}}]{Xavier Porte}
Xavier Porte was born in Manacor, Spain, in 1981. He received the Msc. degree in 2011 and the PhD. degree in 2015, both in Physics, from the University of the Balearic Islands (UIB), Spain. His Msc. and PhD. main research topic was nonlinear dynamics of delay-coupled semiconductor lasers. From 2016 to 2018 he was with the Technical University Berlin, Germany, where he was engaged in research work on external optical coupling of quantum-dot microlasers for nanophotonic applications. He is currently a postdoctoral researcher at the FEMTO-ST Institute in Besan\c{c}on, France, working on photonic spatio-temporal neural networks. His research interests range from nonlinear delay dynamics to nanophotonics and neuromorphic computing. 
\end{IEEEbiography}

\begin{IEEEbiography}[{\includegraphics[width=1in,height=1.25in,clip,keepaspectratio]{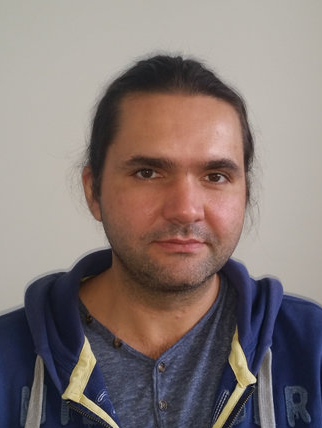}}]{Maxime Jacquot}
Maxime Jacquot received the PhD degree in engineering sciences from the University of Franche-Comté (UFC), France, in 2001. His PhD research was done in the Laboratoire d’Optique PM Duffieux (LOPMD) of Besan\c{c}on and concerned digital microholography and its applications in optical metrology (interferometric holography, phase contrast imaging and microscopy). In 2002, he joined as associate professor the Laboratoire Traitement du Signal et Instrumentation (ex-LTSI, currently Laboratoire Hubert Curien) of Saint Etienne, France. His research interests dealt with spectral interferometry, optical correlator. In 2006, he joined as associate professor the Optics Department of FEMTO-ST Institute, Besan\c{c}on (France). He became full professor at UFC/UBFC in 2018.  His research interests currently include nonlinear delay dynamics and its applications in optical chaos communications and photonic neural networks, digital holography and spatial laser beam shaping with SLM for surface micro- or nano-processing. He received a Habilitation degree (HDR) in 2011 at UFC.

He is deputy director of the FEMTO-ST Optics Dept, and of the International Master PICS - ISITE BFC and Graduate School EIPHI. He is president of the collegium “Master of Excellence for Engineering” at the UFC, in charge of CMI at UFC for a Master of Excellence for Engineering and Research. 
\end{IEEEbiography}

\begin{IEEEbiography}[{\includegraphics[width=1in,height=1.25in,clip,keepaspectratio]{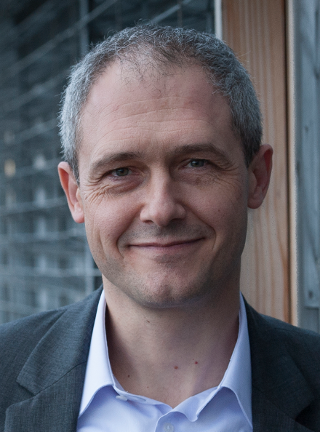}}]{Laurent Larger}
Laurent Larger (SM’11) received the Degree in electronic engineering from the University of Paris XI, Orsay, France, in 1988, the Agrégation degree in applied physics from École Normale Supérieure de Paris-Saclay, Cachan, France, in 1991, the Ph.D. degree in optical engineering, and the Habilitation degree from the University of Franche-Comt\'{e}, Besan\c{c}on, France, in 1997 and 2002, respectively. He was in charge of the International Research Center GTL-CNRS Telecom in Metz, France, a joint laboratory between the French CNRS, Georgia Tech Atlanta, USA, and the University of Franche-Comté, from 2003 to 2006. He became a Full Professor at the University of Franche-Comté in 2005. He is involved in research with the FEMTO-ST Institute, Besan\c{c}on. His research interests have concerned the study of chaos in optical and electronic systems for secure communications, complex behavior in delayed nonlinear dynamics, high spectral purity microwave optoelectronic oscillators, nonlinear dynamics and application of crystalline whispering gallery mode optical resonators, neuromorphic photonic computing exploiting the complexity of nonlinear dynamical transients, and chimeras states in bandpass nonlinear delay dynamics. He is an Honorary Member of the Institut Universitaire de France. He was Deputy Director of the FEMTO-ST Research Institute (750 members), from 2012 to 2016, and in 2017 he became Director of this interdisciplinary research institute in applied physics and engineering sciences. He is Director of the international Graduate School EIPHI at the University Bourgogne-Franche-Comté. He is also President of the non-profit and private foundation FC'INNOV, which business unit FEMTO-Engineering aims to promote industrial innovation sourced by academic research at FEMTO-ST.
\end{IEEEbiography}

\begin{IEEEbiography}[{\includegraphics[width=1in,height=1.25in,clip,keepaspectratio]{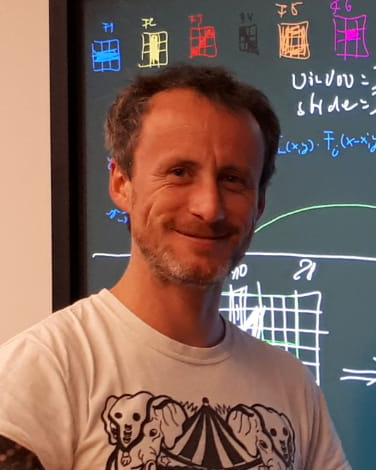}}]{Daniel Brunner}
Daniel Brunner was born in Oedheim, Germany. He studied physics at the “Karlsruhe Institute of Technology”, Karlsruhe, Germany and at Heriot-Watt University, Edinburgh, U.K., where he received the Master’s degree in optoelectronics and lasers and in 2010 a Ph.D. in quantum optics of single quantum dots. He continued as a Postdoctoral Researcher at the Institute for Cross-Disciplinary Physics and Complex Systems, Palma de Mallorca, Spain, holding a Marie Curie Intra European fellowship. Since November 2015, he is a permanent CNRS researcher at the FEMTO-ST, Besan\c{c}on, France.

His research interests are nonlinear photonics with a focus on novel approaches to information processing using quantum systems or nonlinear dynamics. His recent focus lies on the implementation of neural networks and machine learning in photonic systems. He has received several University prizes and the IOP’s 2010 Roy’s prize for his Ph.D. thesis.
\end{IEEEbiography}

\end{document}